\documentclass[%
reprint,
superscriptaddress,
 amsmath,amssymb,
 aps,
]{revtex4-2}

\usepackage{graphicx}
\usepackage{dcolumn}
\usepackage{bm}

\usepackage{babel}

\usepackage{natbib}

\usepackage{datetime2}

\usepackage{letltxmacro}

\LetLtxMacro{\ORIGselectlanguage}{\selectlanguage}
\makeatletter
\DeclareRobustCommand{\selectlanguage}[1]{%
  \@ifundefined{alias@\string#1}
    {\ORIGselectlanguage{#1}}
    {\begingroup\edef\x{\endgroup
       \noexpand\ORIGselectlanguage{\@nameuse{alias@#1}}}\x}%
}
\newcommand{\definelanguagealias}[2]{%
  \@namedef{alias@#1}{#2}%
}
\makeatother

\definelanguagealias{en}{english}

\usepackage{amssymb}

\usepackage{float}

\usepackage{comment}

\usepackage{xcolor}

\begin{document}

\title{Highly-Linear Proximity-Based Bi-SQUID Operating above 4 K}
\author{G. Trupiano} 
\email[Corresponding author: ]{giacomo.trupiano@sns.it}
 \affiliation{NEST, Istituto Nanoscienze-CNR and Scuola Normale Superiore, Piazza S. Silvestro 12, I-56127 Pisa, Italy}

\author{E. Riccardi}
 \affiliation{NEST, Istituto Nanoscienze-CNR and Scuola Normale Superiore, Piazza S. Silvestro 12, I-56127 Pisa, Italy}

\author{C. Puglia}
    \affiliation{INFN Sezione di Pisa, Largo Bruno Pontecorvo 3, I-56127 Pisa, Italy}
    
\author{M. Kiczynski}
\affiliation{Quantum and Nano Technology Group, School of Chemical Engineering, The University of Adelaide, Adelaide, SA 5005, Australia}

\author{A. Gardin}
\affiliation{Quantum and Nano Technology Group, School of Chemical Engineering, The University of Adelaide, Adelaide, SA 5005, Australia}

\author{G. De Simoni}
\email[Corresponding author: ]{giorgio.desimoni@nano.cnr.it}
\affiliation{NEST, Istituto Nanoscienze-CNR and Scuola Normale Superiore, Piazza S. Silvestro 12, I-56127 Pisa, Italy}

\author{G. C. Tettamanzi}
\email[Corresponding author: ]{giuseppe.tettamanzi@adelaide.edu.au}
    \affiliation{Quantum and Nano Technology Group, School of Chemical Engineering, The University of Adelaide, Adelaide, SA 5005, Australia}

\author{F. Giazotto}
\email[Corresponding author: ]{francesco.giazotto@sns.it}
    \affiliation{NEST, Istituto Nanoscienze-CNR and Scuola Normale Superiore, Piazza S. Silvestro 12, I-56127 Pisa, Italy}

\begin{abstract}
We demonstrate a highly linear superconducting quantum interference device (SQUID) amplifier based on a double-loop (Bi-SQUID) architecture incorporating three superconductor-normal metal-superconductor (S-N-S) junctions. Fabricated using niobium-gold technology, the device exhibits robust operation at liquid helium temperatures, with a superconducting transition temperature of 8.5 K. The flux-to-voltage transfer function demonstrates sharp, symmetric, and highly linear behavior at temperatures up to 5 K. Bi-SQUIDs featuring our single-element S-N-S design represent an interesting and original approach to this field, as they demonstrate a numerically estimated spurious-free dynamic range (SFDR) linearity exceeding 60 dB, achieved in a single element, simplifying the requirements in terms of arrays containing hundreds of junctions. These results highlight the potential of proximity-based Bi-SQUIDs for compact, low-noise, and highly linear cryogenic amplifiers in quantum sensing, magnetometry, and biomedical diagnostics.

\end{abstract}



\maketitle

\section{Introduction} \label{sec:introduction}

Superconducting quantum interference devices (SQUIDs) are among the most sensitive detectors of magnetic flux, with broad applications in quantum computing, low-temperature instrumentation, magnetometry, and biomedical diagnostics \cite{barone1982physics, clarke2006squid, Sternickel_2006}. The conventional dc SQUID, composed of a superconducting loop interrupted by two Josephson junctions, produces a periodic voltage response to an applied magnetic flux and operates as a high-sensitivity, low-noise transducer. Despite its sensitivity, the dc SQUID suffers from an intrinsically non-linear flux-to-voltage transfer function \cite{10.1063/1.1377043}, which limits its dynamic range and generates harmonic distortion when used in analog amplification applications. To overcome these limitations, superconducting double-loop interferometers (Bi-SQUIDs) were proposed as a modification of the dc SQUID to improve linearity~\cite{kornevBiSQUIDNovelLinearization2009,kornevSignalNoiseCharacteristics2014,kornevHighInductanceBiSQUID2017,kornevHighLinearityBiSQUIDDesign2018,kornevBiSQUIDDesignApplications2020}. The Bi-SQUID integrates a second superconducting loop into the device layout and adds a third Josephson junction, enabling destructive interference of nonlinear components in the transfer function. The resulting voltage-flux characteristics exhibit improved linearity, which is crucial for precision analog applications such as analog-to-digital conversion and wide-band flux detection.

Most prior implementations of Bi-SQUID have relied on superconductor-insulator-superconductor (S-I-S) tunnel junctions, where linearity is often further improved by constructing large series arrays of identical Bi-SQUID elements \cite{5672560}. These arrays average out residual non-linearity and improve device gain, enabling spurious-free dynamic ranges (SFDR) exceeding 100 dB in optimized systems ~\cite{kornevBiSQUIDNovelLinearization2009,kornevSignalNoiseCharacteristics2014,kornevHighInductanceBiSQUID2017,kornevHighLinearityBiSQUIDDesign2018,kornevBiSQUIDDesignApplications2020}. However, for some applications, such arrays introduce unwanted fabrication complexity 
and potential non-uniformity across elements, which can degrade performance in practical applications. Therefore, in some instances, superconductor-normal metal-superconductor (S-N-S) junctions represent a compelling alternative to conventional S-I-S systems~\cite{PhysRevApplied.18.014073}. In S-N-S structures, the proximity effect~\cite{PhysRevLett.25.507} creates Andreev bound states in the normal metal region, enabling non-dissipative super-current transport. The key advantage of S-N-S junctions lies in the independent control of the critical current and capacitance. While S-I-S junctions require careful optimization of junction area and tunnel barrier opacity, parameters that simultaneously determine both critical current and capacitance, S-N-S junctions achieve critical current tuning purely through geometric parameters (length, width, overlap area) while maintaining inherently negligible capacitance. This parameter decoupling eliminates capacitance-related constraints and enables precise engineering of critical current~\cite{likharevSuperconductingWeakLinks1979} ratios and current phase relations \cite{golubovCurrentphaseRelationJosephson2004} through lithographic control rather than complex tunnel barrier fabrication.

In contrast to array-based approaches, we demonstrate that a single proximity-based Bi-SQUID element can achieve high linearity without requiring large-scale integration. 
Here, we present the design, fabrication, and experimental characterization of a single Bi-SQUID amplifier incorporating three long \cite{likharevSuperconductingWeakLinks1979} diffusive S-N-S junctions. The device exhibits a sharp and symmetric voltage response with exceptional linearity over a broad flux range, outperforming standard dc SQUIDs with robust operation up to 5 K. These results highlight the potential of S-N-S Bi-SQUIDs for compact, low-noise, and highly linear cryogenic amplifiers, which offer superior design flexibility combined with simplified fabrication processes for next-generation superconducting transducers.

\section{Results and Discussion} \label{sec:results}

\begin{figure}[t!]
\includegraphics[width=\linewidth]{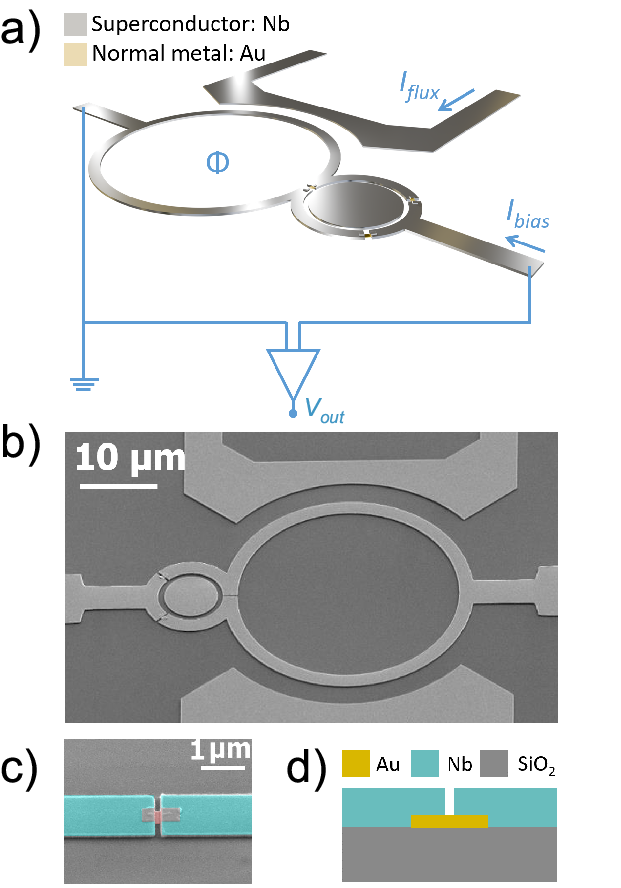}
\caption{\label{fig:fig1} Device scheme and SEM image.
a) Circuit diagram of the Bi-SQUID. A flux current $I_{flux}$ injected into the external flux line generates magnetic flux $\Phi$ that threads the large loop, but is screened from the small loop by a superconducting layer. Three superconductor-normal metal-superconductor (S-N-S) junctions interrupt the small loop, with one junction shared between both loops. A bias current flows through the device structure while a voltage is measured across it.
b) SEM image of the Bi-SQUID viewed from above. The large loop has an internal diameter of 28 $\mu$m, while the small loop has an internal diameter of 8 $\mu$m.
c) False-colour SEM image of an S-N-S junction viewed from above. Note that this junction is fabricated using the same process as those used in the Bi-SQUID, but it is not one of the junctions shown in panel b).
d) Schematic of the layer structure of the S-N-S junction: a gold (Au) layer with a thickness of 30 nm, length of 800 nm, and width of 300 nm is deposited on SiO$_2$. A 120 nm thick niobium (Nb) layer is deposited on top. This layer is then etched to create a 250 nm gap, which forms the S-N-S junction.}
\end{figure}

We fabricated and characterized a niobium-gold Bi-SQUID amplifier optimized for high linearity at liquid helium temperatures. As shown in Figure~\ref{fig:fig1}, the device consists of a large superconducting loop (28 $\mu$m internal diameter) incorporating a smaller loop (8 $\mu$m internal diameter), which is interrupted by three superconductor-normal metal-superconductor (S-N-S) junctions. The design features the typical layout of a Bi-SQUID, in which one of the S-N-S junctions is shared between the two loops, enabling flux-to-voltage conversion with enhanced linearity. A schematic of the circuit and scanning electron microscope (SEM) images of the device are shown in Figure~\ref{fig:fig1}. The S-N-S junctions are manufactured by depositing a 30 nm thick gold layer (Au), 300 nm wide and 800 nm long on an industrially grown 300 nm thick silicon oxide substrate (SiO$_2$), followed by a 120 nm thick niobium layer (Nb), which was patterned to create 250 nm wide gaps.

\begin{figure}[t!]
\includegraphics[width=\linewidth]{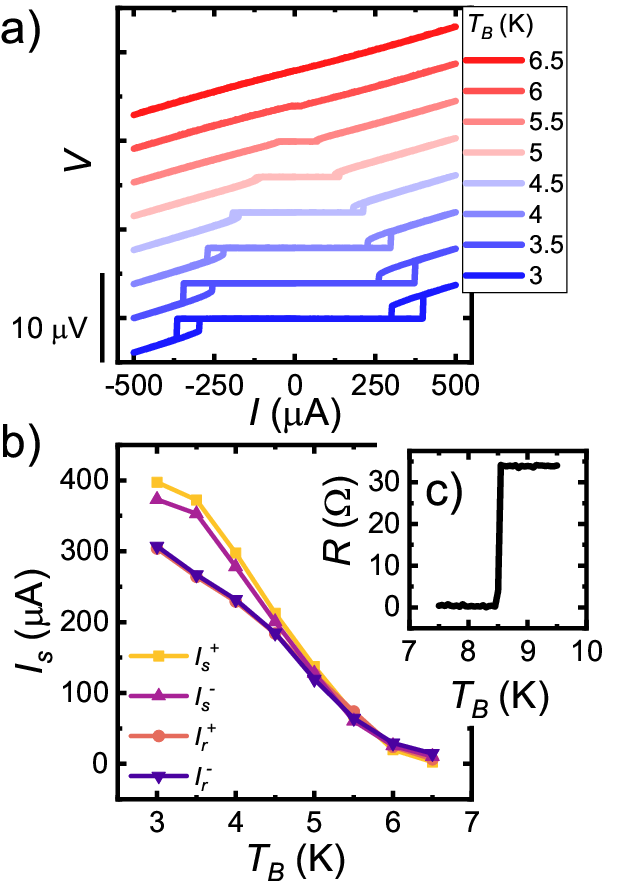}
\caption{\label{fig:fig2} Current-voltage ($IV$) curves, switching current ($I_s$) and resistance ($R$) as a function of bath temperature ($T_B$). 
a) Current-voltage ($IV$) curves measured at various bath temperatures ($T_B$). Below 5 K, the curves exhibit hysteretic behaviour with distinct switching and retrapping currents, typical of S-N-S junctions. The pronounced asymmetry between positive and negative bias directions results from flux screening effects, as quantified later in the text.
b) Switching current ($I_s$) and retrapping current ($I_r$) as a function of bath temperature ($T_B$). The switching current exhibits a strong temperature dependence, characteristic of long diffusive S-N-S junctions.
c) Resistance versus bath temperature of the Bi-SQUID. The data show a single superconducting transition at approximately $T_c=8.5$ K, indicating high junction transparency and proximity effect in the S-N-S junctions, which exhibit the same critical temperature as the niobium leads.}
\end{figure}

Figure~\ref{fig:fig2} summarizes the DC transport properties of Bi-SQUID. Current-voltage characteristics ($IV$) measured below 5 K exhibit evident hysteresis, with distinct switching and retrapping currents ($I_s$ and $I_r$), as typically observed in S-N-S junctions \cite{PhysRevB.63.064502, PhysRevLett.101.067002}. This hysteresis is thermal in origin and becomes more pronounced at lower temperatures. A notable asymmetry between the positive ($I_s^+$) and negative ($I_s^-$) critical currents emerges and intensifies as the temperature decreases, as shown in in Fig.~\ref{fig:fig2}(b). This effect arises from flux screening in the superconducting loops, as further discussed later in the text.
The switching current $I_s$ exhibits strong temperature dependence, decreasing monotonically with increasing bath temperature ($T_B$), consistent with the thermal suppression of the proximity-induced minigap in the normal metal. In contrast, the retrapping current $I_r$ shows a weaker temperature dependence. Resistance measurements as a function of $T_B$ reveal a sharp superconducting transition at $T_c \simeq 8.5$ K, indicating high interface transparency and a uniform proximity effect across all S-N-S junctions. 

The three S-N-S junctions operate in the diffusive transport regime and satisfy the long-junction limit condition \cite{likharevSuperconductingWeakLinks1979}. The characteristic energy scale for phase coherence, Thouless energy, is given by $E_{Th} = \hbar D / l^2 \simeq 0.1$ meV, which is significantly smaller than the superconducting energy gap of the niobium leads, $\Delta_{Nb} = 1.764 k_B T_c\simeq 1.3$ meV. Here, $D = 0.01$ m$^2$/s is the diffusion coefficient of gold \cite{PhysRevB.77.165408}, $k_B$ is the Boltzmann constant, and $l=250$ nm is the length of the normal-metal weak link. In this long-junction limit ($E_{Th} \ll \Delta_{Nb}$), the supercurrent in the S-N-S junctions is governed by a sinusoidal current-phase relation.


The Resistively and Capacitively Shunted Junction (RCSJ) model is the standard framework for describing Josephson tunnel junctions and is widely used for S-I-S devices. However, in our case, the junctions are long and diffusive S-N-S weak links, which exhibit a sinusoidal current–phase relationship \cite{golubovCurrentphaseRelationJosephson2004} and negligible capacitance. Under these conditions, the RCSJ model reduces to the simpler Resistively Shunted Junction (RSJ) limit. We therefore describe the current–flux characteristics of the Bi-SQUID within the RSJ model. Although the Bi-SQUID includes an additional superconducting loop that modifies the total enclosed magnetic flux, the general RCSJ formalism remains applicable to such devices~\cite{clarke2006squid,Kong_2024}. In our case, this formalism reduces to the RSJ limit, but must incorporate flux quantization conditions for both loops. The resulting dynamics differ qualitatively from those of conventional dc SQUIDs and are described by the following set of coupled equations:

\begin{align}
I_{bias} &= I_0\left[(1 + \alpha)\sin(\delta_A) + (1 - \alpha)\sin(\delta_B)\right], \label{eq:current_balance} \\
\delta_C &= \frac{2\pi \Phi}{\Phi_0} - \pi \beta_1 j_1 = \frac{2\pi \Phi}{\Phi_0} -\pi \beta_1 j_2 \nonumber\\
& \quad + 2\pi \beta_1 \alpha_3 \sin(\delta_C), \label{eq:flux_loop1} \\
\delta_C &= \delta_B - \delta_A - \pi \beta_2 j_2 + \frac{2\pi \Phi_2}{\Phi_0}. \label{eq:flux_loop2}
\end{align}
Here, $\Phi_0 = h/2e \simeq 2.067 \times 10^{-15}$ Wb is the magnetic flux quantum, where $h$ is the Planck constant and $e$ is the elementary charge, while $\Phi$ is the magnetic flux through the large loop. The parameter $\alpha = \frac{I_A - I_B}{I_A + I_B}$ quantifies the asymmetry between the critical currents $I_A$ and $I_B$ of junctions A and B, which are located in the smaller loop. The quantity $I_0 = \frac{I_A + I_B}{2}$ corresponds to half the total critical current of the interferometer, which is the average critical current of junctions A and B. The parameter $\alpha_3 = \frac{I_C}{I_0}$ represents the normalized critical current of the third junction, which is shared between the two loops. The phase differences across the three weak links are indicated by $\delta_A$, $\delta_B$, and $\delta_C$, respectively. The circulating currents in loops 1 and 2 (normalized by $I_0$) are represented by $j_1$ and $j_2$. Finally, the screening parameters $\beta_1$ and $\beta_2$ account for the inductances of the two loops and determine the strength of the flux-current coupling. In our device, the contribution of magnetic flux threading the small loop can be neglected due to the presence of a superconducting niobium shield, which effectively reduces its area. As a result, in Eq.~\ref{eq:flux_loop2}, we can omit the term $\pi \beta_2 j_2$ and the term $\frac{2\pi \Phi_2}{\Phi_0}$, where $\Phi_2$ is the magnetic flux through the small loop.

\begin{figure}[t!]
\includegraphics[width=\linewidth]{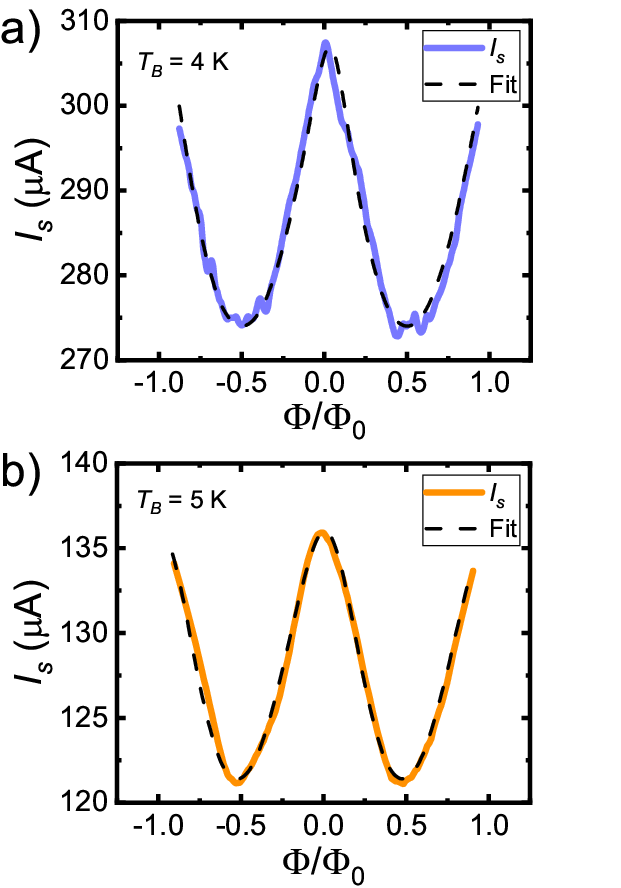}
\caption{\label{fig:fig3} Switching current ($I_s$) vs magnetic flux applied ($\Phi$). a) Measured switching current ($I_s$) with the superimposed fitted curve of the model obtained from Eq.~\eqref{eq:current_balance}, Eq.~\eqref{eq:flux_loop1}, and Eq.~\eqref{eq:flux_loop2}  at $T_B = 4$ K. The magnetic flux is applied by current-biasing the external flux line shown in Figure~\ref{fig:fig1}(a) and Figure~\ref{fig:fig1}(b). b) Measured switching current ($I_s$) with superimposed the fitted curve at $T_B = 5$ K.  
}
\end{figure}

Figure~\ref{fig:fig3} shows the response of the device to an externally applied magnetic flux. We measured the switching current as a function of the applied flux $\Phi$ through the large loop using the on-chip flux line. The resulting $I_s(\Phi)$ characteristics exhibit periodic SQUID oscillations with a period of one flux quantum $\Phi_0$. 
We fitted the measured switching current using the system of equations introduced above to extract the device parameters. The fitting procedure converges even when the terms $\pi \beta_2 j_2$ and $\frac{2\pi \Phi_2}{\Phi_0}$ of Eq.~\ref{eq:flux_loop2} are neglected, confirming that the niobium shield effectively screens the small loop. The absence of flux threading the smaller loop is a desirable feature, as it improves the linearity of the device. The fitted switching current curves at 4 K and 5 K are reported in Figure~\ref{fig:fig3}. The extracted parameters are $\alpha = 0.893 \pm 0.001$, $\alpha_3 = 0.241 \pm 0.005$, and $\beta_1 = 1.909 \pm 0.007$ at 4 K; and $\alpha = 0.893 \pm 0.001$, $\alpha_3 = 0.113 \pm 0.005$, and $\beta_1 = 1.813 \pm 0.006$ at 5 K. 

The large value of $\alpha$ confirms the asymmetry of the two junctions in the small loop. This explains the reduced visibility of the interference pattern, which originates from the uncertainty of nanofabrication: the critical current of long diffusive S–N–S junctions depends exponentially on their length \cite{likharevSuperconductingWeakLinks1979}. The parameter $\alpha_3$ indicates that the critical current of the third junction is about one quarter of the average value of the other two, suggesting that this junction is longer than expected. Its pronounced decrease with temperature reflects the fact that longer junctions show a stronger temperature dependence of their critical current. Finally, the product $\alpha \beta_1$ determines the relative shift in the magnetic field between the positive and negative branches of the critical current. Perfect alignment requires $\alpha \beta_1 \ll 1$. In our case, $\alpha \beta_1 \simeq 1.7$, resulting in a significant shift. This accounts for the asymmetry of the IV curves at fixed magnetic flux, as shown in Figure~\ref{fig:fig2}. Finally, the small decrease in $\beta_1$ in temperature can be explained by a decrease in the kinetic inductance of the niobium.

\begin{figure}[t!]
\includegraphics[width=\linewidth]{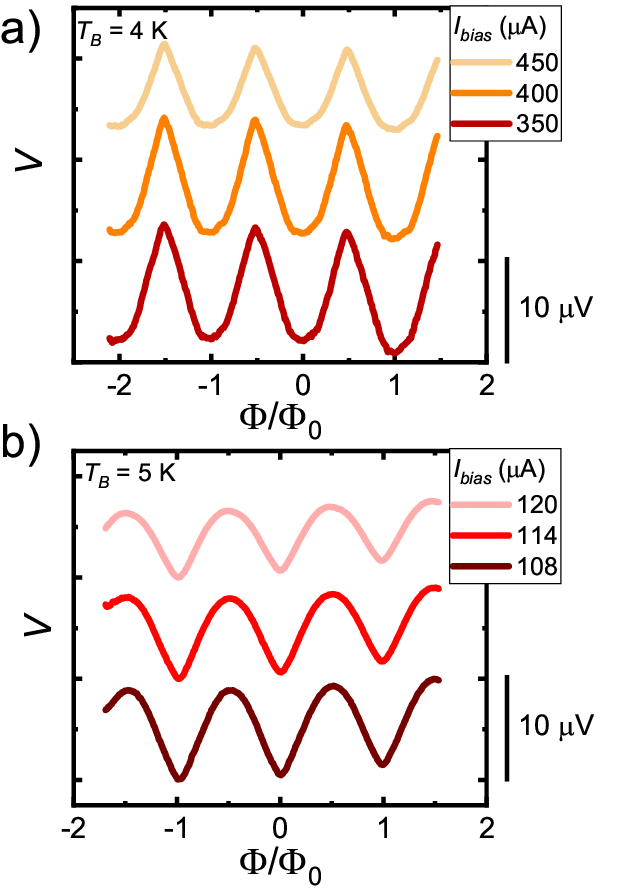}
\caption{\label{fig:fig4} Voltage ($V$) as a function of applied magnetic flux $\Phi$ for different bias current applied ($I_{bias}$).
a) Four-wire lock-in measurements of voltage ($V$) versus applied magnetic flux ($\Phi$) at $T_B = 4$ K for different bias currents ($I_{bias}$). The device is biased with a periodic ramp signal generated by an arbitrary waveform generator at a frequency of 17 Hz. The ramp modulation sweeps from 0 to the maximum values indicated in the legend, effectively removing the contribution of retrapping current from the voltage measured by the lock-in amplifier. The curves exhibit the sharp shape and highly linear flux-to-voltage transfer function characteristic of a Bi-SQUID.
b) Voltage-flux characteristics measured under identical conditions at $T_B = 5$ K. The transfer function maintains its sharp, linear behavior at higher temperatures, demonstrating stable Bi-SQUID performance across the operating temperature range.}
\end{figure}

Figure~\ref{fig:fig4} presents the flux-to-voltage transfer characteristics of the Bi-SQUID, measured using low-frequency (17 Hz) lock-in detection in a four-wire configuration. The device was biased with a periodic ramp signal generated by an arbitrary waveform generator, sweeping from 0 to the maximum values indicated in the legend. This ramp modulation effectively suppresses the contribution of the retrapping current by ensuring that the voltage is sampled only during the forward sweep, thereby isolating the switching branch of the $I$-$V$ characteristic. The resulting $V(\Phi)$ curves display sharp, symmetric shapes and a nearly linear transfer function over a broad flux range at $T_B = 4$ K. A similar behavior is observed at $T_B = 5$ K, confirming that the high-linearity performance of the device is preserved in the relevant temperature range.

\begin{figure}[t!]
\includegraphics[width=\linewidth]{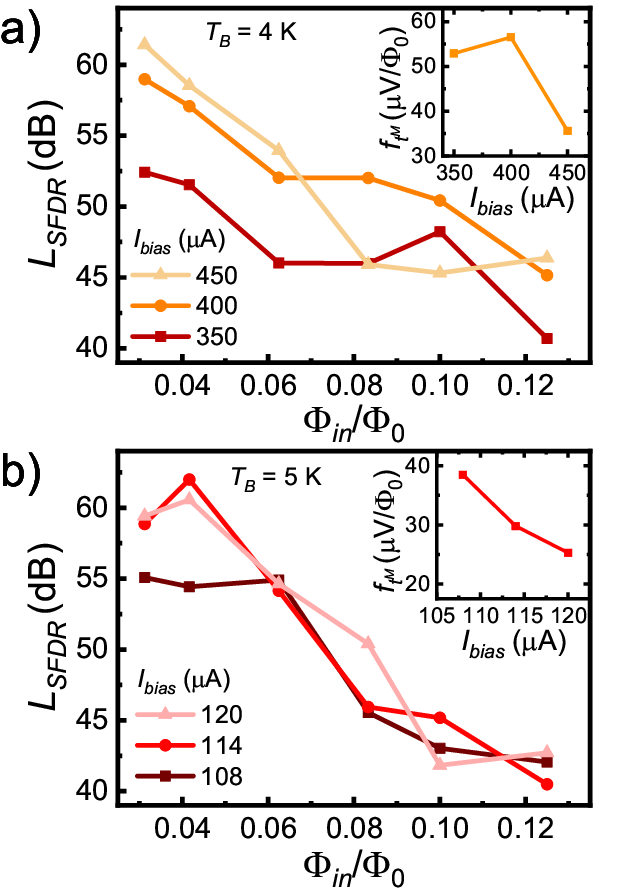}
\caption{\label{fig:fig5} Figures of merit: Maximum spurious-free dynamic range linearity ($L_{SFDR}$) and flux-to-voltage transfer function ($f_{t^M}$) for different temperatures, bias currents, and input flux amplitudes.
a) Maximum spurious-free dynamic range linearity ($L_{SFDR}$) as a function of input flux amplitude ($\Phi_{in}$) for different bias currents at $T_B = 4$ K. The linearity is extracted from the voltage-flux response of the Bi-SQUID by numerically applying a sinusoidal flux modulation of amplitude $\Phi_{in}$ to the measured transfer characteristics from Figure~\ref{fig:fig4}, then analyzing the output signal harmonics using Fast Fourier Transform (FFT). The inset shows the maximum value of the flux-to-voltage transfer function ($f_{t^M}$) as a function of the bias current at 4 K.
b) Spurious-free dynamic range estimation under identical conditions at $T_B = 5$ K. The comparable performance at both temperatures demonstrates the robust linearity of the Bi-SQUID flux-to-voltage conversion across the operating range. The inset displays $f_{t^M}$ versus bias current at 5 K.}
\end{figure}

One of the conventional figures of merit of a SQUID is represented by the maximum flux-to-voltage transfer function $f_{t^M} = \partial V / \partial \Phi$, which is shown in the insets of Figure~\ref{fig:fig5}. We obtained values on the order of several tens of $\mu$V/$\Phi_0$, which are significantly lower than those reported for other Bi-SQUIDs in the literature \cite{PhysRevApplied.18.014073}. This reduced performance is primarily attributed to the low visibility of the SQUID oscillations caused by the high asymmetry between the two S-N-S junctions in the small loop. This is a consequence of the physical properties of the S-N-S junctions in the long-diffusive regime, in which the critical current scales exponentially with the junction length. For this reason, even minor fabrication imperfections can generate highly asymmetric junctions. In future work, we plan to address this limitation by symmetrizing the critical currents of the two junctions, either by electrostatic gating (exploiting gate-controlled superconductivity) \cite{ruf2024gate,desimoniJosephsonFieldEffectTransistors2019, desimoniGateControlCurrent2021a,bours2020unveiling} or by local heating of one junction using quasiparticle injection \cite{mccaughan_superconducting-nanowire_2014, baghdadi_multilayered_2020, buzzi_nanocryotron_2023,savin2004cold,PhysRevApplied.23.014046,tirelli2008manipulation, trupiano2025thermallymodulatedsinistrasconductance}.

An estimate of $f_{t^{M}}$ provides an upper bound for the flux noise sensitivity of the S-N-S Bi-SQUID. Previous studies on tunnel-based Bi-SQUIDs indicate that their output voltage noise is comparable to, and typically about twice as large as, that of equivalent dc SQUIDs. This is primarily attributed to the additional noise from the third junction and the nonlinear flux-to-phase conversion used for linearity enhancement. We can reasonably extend these considerations to our proximity-based Bi-SQUID. Under this assumption, the dominant noise source in our setup is the room-temperature voltage preamplifier, which introduces a voltage noise spectral density no less than 6.5 nV/$\sqrt{\mathrm{Hz}}$. This value is roughly an order of magnitude higher than that of S-N-S dc SQUIDs \cite{10.1063/1.4817013}. Taking the ratio of this voltage noise to $f_{t^{M}}$, we estimate an upper limit for our device flux noise sensitivity to be approximately $0.1$ m$\Phi_0/\sqrt{\mathrm{Hz}}$, which is mainly limited by the low $f_{t^M}$ values we discussed previously.

To quantify the linearity of the Bi-SQUID amplifier, we computed the spurious-free dynamic range ($L_{SFDR}$), defined as:
\begin{align}
L_{\text{SFDR}} = 20\log\left(\frac{A_{1}}{A_{\text{spur}}}\right),
\end{align}
where $A_1$ is the main harmonic and $A_{\text{spur}}$ is the highest spurious harmonic, digitally modulating the measured $V(\Phi)$ curves with a sinusoidal flux input and analyzing the harmonic distortion by the fast Fourier Transform \cite{PhysRevApplied.18.014073}. In Figure~\ref{fig:fig5} we show that $L_{SFDR}$ reaches its maximum for lower input amplitudes, as expected. The results yielded linearity values within the range of 40-60 dB. However, these figures should be considered to be upper limits. This consideration arises from the finite sampling rate employed in numerical estimation, which may lead to an underestimation of the amplitude of higher-order harmonics. At both 4 K and 5 K, the device achieves linearity values exceeding those typical of standard dc SQUIDs, confirming the advantage of the Bi-SQUID geometry for linear flux detection. The consistent performance across temperatures highlights the device potential for low-noise amplification in cryogenic environments.
\section{Conclusion} \label{sec:conclusion}

We have demonstrated a niobium-gold S-N-S Bi-SQUID amplifier with exceptional linearity and robust performance at liquid helium temperatures. The flux-to-voltage transfer function remains sharp, symmetric, and highly linear over a broad flux range, with stable operation up to 5 K. A numerical analysis of the spurious-free dynamic range (SFDR) reveals linearity exceeding 60 dB, achieved with a single Bi-SQUID element.

This represents a significant advancement, as S-N-S junctions enable simpler architectures while maintaining high performance. These results establish S-N-S Bi-SQUIDs as a competitive platform for applications requiring high linearity and low noise, including quantum readout, precision magnetometry, and cryogenic analog-to-digital conversion. 

Future improvements could focus on symmetrizing the critical currents of the two junctions in the small loop, using electrostatic gating (to exploit gate-controlled superconductivity) or localized quasiparticle injection via a tunnel junction. Such strategies offer a promising route to enhanced flux visibility and noise performance, further advancing the capabilities of proximity-based Bi-SQUID cryogenic amplifiers.

\section*{Acknowledgements} \label{sec:acknowledgements}

We acknowledge the University of Adelaide CAS funding body project AE230971 and the PNRR MUR project PE0000023-NQSTI for partial
financial support.

\appendix

\section{Device nanofabrication} \label{sec:appendix1}
 Ti-Au wires are defined on an Si-SiO$_2$ substrate using electron-beam lithography (EBL) followed by thermal evaporation. The gold is then treated with Argon plasma cleaning in the load lock chamber of the sputtering system. This process enables the removal of a few nanometers of metal, resulting in a clean interface. Subsequently, the sample is introduced into the ultra-high vacuum chamber (10$^{-9}$ mbar) where a 120 nm layer of niobium is deposited via DC sputtering (150 W,  5$\cdot 10^{-4}$ mbar).
A 25-nm-thick aluminum hard mask is then fabricated on top of the niobium film using EBL lithography and thermal evaporation. This mask is used to create superconducting loops and junctions, 250 nm long, through Reactive Ion Etching (RIE) with a mixture of CF$_4$, Ar, and O$_2$ (10:1:1). The hard mask is finally removed using wet aluminium etching. 
\section{Cryogenic electrical characterization} \label{sec:appendix2}
The electrical measurements of the interferometer are performed using a four-wire technique in an Oxford ICE 3 K dry cryostat. All measurement leads are filtered using $\pi$-filters at room temperature and additional RC low-pass filters ($f_c\simeq 1$ kHz) in the cryogenic stage to minimize noise pickup from high frequencies. The cryostat maintains a base temperature of 2.8 K with temperature stability better than $\pm$10 mK during the measurements. Temperature control was provided by a Lake Shore 340 temperature controller, which used a calibrated Cernox thermometer positioned near the sample stage.

Current-voltage measurements are performed by applying a current bias through a Yokogawa GS200 voltage source in series with a bias resistor. The bias current is determined from the voltage drop across the calibrated bias resistor. The voltage drop across the device is measured using an HP 34401A digital multimeter, which has been amplified by a low-noise Femto DLPVA voltage amplifier with a gain of 80 dB, a voltage noise spectral density of approximately 6.5 nV/$\sqrt{\mathrm{Hz}}$, and a bandwidth of 1 kHz. 

The voltage-flux measurements are performed via a low-frequency (17 Hz) lock-in technique to isolate the switching branch of the I-V characteristic. An Agilent 33220A arbitrary wave generator generates a voltage ramp with a 17 Hz frequency applied across a bias resistor, creating a corresponding current ramp through the device. The subsequent modulation of the current ramp ranges from 0 to a value higher than the switching current of the device, effectively removing the contribution of retrapping current from the voltage subsequently measured by an NF LI 5640 lock-in amplifier with an integration time of 3 seconds. The external magnetic flux is applied using an on-chip flux line fabricated along with the Bi-SQUID structure. The flux line current is controlled by a second Yokogawa GS200 source, with the flux-current conversion factor calibrated from the periodicity of the SQUID oscillations (one flux quantum $\Phi_0$ per period).


\begin{thebibliography}{30}%
\makeatletter
\providecommand \@ifxundefined [1]{%
 \@ifx{#1\undefined}
}%
\providecommand \@ifnum [1]{%
 \ifnum #1\expandafter \@firstoftwo
 \else \expandafter \@secondoftwo
 \fi
}%
\providecommand \@ifx [1]{%
 \ifx #1\expandafter \@firstoftwo
 \else \expandafter \@secondoftwo
 \fi
}%
\providecommand \natexlab [1]{#1}%
\providecommand \enquote  [1]{``#1''}%
\providecommand \bibnamefont  [1]{#1}%
\providecommand \bibfnamefont [1]{#1}%
\providecommand \citenamefont [1]{#1}%
\providecommand \href@noop [0]{\@secondoftwo}%
\providecommand \href [0]{\begingroup \@sanitize@url \@href}%
\providecommand \@href[1]{\@@startlink{#1}\@@href}%
\providecommand \@@href[1]{\endgroup#1\@@endlink}%
\providecommand \@sanitize@url [0]{\catcode `\\12\catcode `\$12\catcode `\&12\catcode `\#12\catcode `\^12\catcode `\_12\catcode `\%12\relax}%
\providecommand \@@startlink[1]{}%
\providecommand \@@endlink[0]{}%
\providecommand \url  [0]{\begingroup\@sanitize@url \@url }%
\providecommand \@url [1]{\endgroup\@href {#1}{\urlprefix }}%
\providecommand \urlprefix  [0]{URL }%
\providecommand \Eprint [0]{\href }%
\providecommand \doibase [0]{https://doi.org/}%
\providecommand \selectlanguage [0]{\@gobble}%
\providecommand \bibinfo  [0]{\@secondoftwo}%
\providecommand \bibfield  [0]{\@secondoftwo}%
\providecommand \translation [1]{[#1]}%
\providecommand \BibitemOpen [0]{}%
\providecommand \bibitemStop [0]{}%
\providecommand \bibitemNoStop [0]{.\EOS\space}%
\providecommand \EOS [0]{\spacefactor3000\relax}%
\providecommand \BibitemShut  [1]{\csname bibitem#1\endcsname}%
\let\auto@bib@innerbib\@empty
\bibitem [{\citenamefont {Barone}\ and\ \citenamefont {Paterno}(1982)}]{barone1982physics}%
  \BibitemOpen
  \bibfield  {author} {\bibinfo {author} {\bibfnamefont {A.}~\bibnamefont {Barone}}\ and\ \bibinfo {author} {\bibfnamefont {G.}~\bibnamefont {Paterno}},\ }\href@noop {} {\emph {\bibinfo {title} {Physics and applications of the josephson effect}}},\ A {Wiley}-interscience publication\ (\bibinfo  {publisher} {Wiley},\ \bibinfo {year} {1982})\BibitemShut {NoStop}%
\bibitem [{\citenamefont {Clarke}\ and\ \citenamefont {Braginski}(2006)}]{clarke2006squid}%
  \BibitemOpen
  \bibfield  {author} {\bibinfo {author} {\bibfnamefont {J.}~\bibnamefont {Clarke}}\ and\ \bibinfo {author} {\bibfnamefont {A.}~\bibnamefont {Braginski}},\ }\href {https://books.google.it/books?id=buntRdQD7I8C} {\emph {\bibinfo {title} {The {SQUID} handbook: {Applications} of squids and {SQUID} systems}}}\ (\bibinfo  {publisher} {Wiley},\ \bibinfo {year} {2006})\BibitemShut {NoStop}%
\bibitem [{\citenamefont {Sternickel}\ and\ \citenamefont {Braginski}(2006)}]{Sternickel_2006}%
  \BibitemOpen
  \bibfield  {author} {\bibinfo {author} {\bibfnamefont {K.}~\bibnamefont {Sternickel}}\ and\ \bibinfo {author} {\bibfnamefont {A.~I.}\ \bibnamefont {Braginski}},\ }\bibfield  {title} {\bibinfo {title} {Biomagnetism using {SQUIDs}: status and perspectives},\ }\href {https://doi.org/10.1088/0953-2048/19/3/024} {\bibfield  {journal} {\bibinfo  {journal} {Supercond. Sci. Technol.}\ }\textbf {\bibinfo {volume} {19}},\ \bibinfo {pages} {S160} (\bibinfo {year} {2006})}\BibitemShut {NoStop}%
\bibitem [{\citenamefont {Mück}\ and\ \citenamefont {Clarke}(2001)}]{10.1063/1.1377043}%
  \BibitemOpen
  \bibfield  {author} {\bibinfo {author} {\bibfnamefont {M.}~\bibnamefont {Mück}}\ and\ \bibinfo {author} {\bibfnamefont {J.}~\bibnamefont {Clarke}},\ }\bibfield  {title} {\bibinfo {title} {Harmonic distortion and intermodulation products in the microstrip amplifier based on a superconducting quantum interference device},\ }\href {https://doi.org/10.1063/1.1377043} {\bibfield  {journal} {\bibinfo  {journal} {Appl. Phys. Lett.}\ }\textbf {\bibinfo {volume} {78}},\ \bibinfo {pages} {3666} (\bibinfo {year} {2001})}\BibitemShut {NoStop}%
\bibitem [{\citenamefont {Kornev}\ \emph {et~al.}(2009)\citenamefont {Kornev}, \citenamefont {Soloviev}, \citenamefont {Klenov},\ and\ \citenamefont {Mukhanov}}]{kornevBiSQUIDNovelLinearization2009}%
  \BibitemOpen
  \bibfield  {author} {\bibinfo {author} {\bibfnamefont {V.~K.}\ \bibnamefont {Kornev}}, \bibinfo {author} {\bibfnamefont {I.~I.}\ \bibnamefont {Soloviev}}, \bibinfo {author} {\bibfnamefont {N.~V.}\ \bibnamefont {Klenov}},\ and\ \bibinfo {author} {\bibfnamefont {O.~A.}\ \bibnamefont {Mukhanov}},\ }\bibfield  {title} {{\selectlanguage {en}\bibinfo {title} {Bi-{SQUID}: a novel linearization method for dc {SQUID} voltage response}},\ }\href {https://doi.org/10.1088/0953-2048/22/11/114011} {\bibfield  {journal} {\bibinfo  {journal} {Supercond. Sci. Technol.}\ }\textbf {\bibinfo {volume} {22}},\ \bibinfo {pages} {114011} (\bibinfo {year} {2009})}\BibitemShut {NoStop}%
\bibitem [{\citenamefont {Kornev}\ \emph {et~al.}(2014)\citenamefont {Kornev}, \citenamefont {Sharafiev}, \citenamefont {Soloviev},\ and\ \citenamefont {Mukhanov}}]{kornevSignalNoiseCharacteristics2014}%
  \BibitemOpen
  \bibfield  {author} {\bibinfo {author} {\bibfnamefont {V.~K.}\ \bibnamefont {Kornev}}, \bibinfo {author} {\bibfnamefont {A.~V.}\ \bibnamefont {Sharafiev}}, \bibinfo {author} {\bibfnamefont {I.~I.}\ \bibnamefont {Soloviev}},\ and\ \bibinfo {author} {\bibfnamefont {O.~A.}\ \bibnamefont {Mukhanov}},\ }\bibfield  {title} {{\selectlanguage {en}\bibinfo {title} {Signal and noise characteristics of bi-{SQUID}}},\ }\href {https://doi.org/10.1088/0953-2048/27/11/115009} {\bibfield  {journal} {\bibinfo  {journal} {Supercond. Sci. Technol.}\ }\textbf {\bibinfo {volume} {27}},\ \bibinfo {pages} {115009} (\bibinfo {year} {2014})}\BibitemShut {NoStop}%
\bibitem [{\citenamefont {Kornev}\ \emph {et~al.}(2017)\citenamefont {Kornev}, \citenamefont {Kolotinskiy}, \citenamefont {Bazulin},\ and\ \citenamefont {Mukhanov}}]{kornevHighInductanceBiSQUID2017}%
  \BibitemOpen
  \bibfield  {author} {\bibinfo {author} {\bibfnamefont {V.~K.}\ \bibnamefont {Kornev}}, \bibinfo {author} {\bibfnamefont {N.~V.}\ \bibnamefont {Kolotinskiy}}, \bibinfo {author} {\bibfnamefont {D.~E.}\ \bibnamefont {Bazulin}},\ and\ \bibinfo {author} {\bibfnamefont {O.~A.}\ \bibnamefont {Mukhanov}},\ }\bibfield  {title} {{\selectlanguage {en}\bibinfo {title} {High-{Inductance} {Bi}-{SQUID}}},\ }\href {https://doi.org/10.1109/TASC.2016.2631427} {\bibfield  {journal} {\bibinfo  {journal} {IEEE Trans. Appl. Supercond.}\ }\textbf {\bibinfo {volume} {27}},\ \bibinfo {pages} {1} (\bibinfo {year} {2017})}\BibitemShut {NoStop}%
\bibitem [{\citenamefont {Kornev}\ \emph {et~al.}(2018)\citenamefont {Kornev}, \citenamefont {Kolotinskiy}, \citenamefont {Bazulin},\ and\ \citenamefont {Mukhanov}}]{kornevHighLinearityBiSQUIDDesign2018}%
  \BibitemOpen
  \bibfield  {author} {\bibinfo {author} {\bibfnamefont {V.~K.}\ \bibnamefont {Kornev}}, \bibinfo {author} {\bibfnamefont {N.~V.}\ \bibnamefont {Kolotinskiy}}, \bibinfo {author} {\bibfnamefont {D.~E.}\ \bibnamefont {Bazulin}},\ and\ \bibinfo {author} {\bibfnamefont {O.~A.}\ \bibnamefont {Mukhanov}},\ }\bibfield  {title} {{\selectlanguage {en}\bibinfo {title} {High-{Linearity} {Bi}-{SQUID}: {Design} {Map}}},\ }\href {https://doi.org/10.1109/TASC.2018.2827982} {\bibfield  {journal} {\bibinfo  {journal} {IEEE Trans. Appl. Supercond.}\ }\textbf {\bibinfo {volume} {28}},\ \bibinfo {pages} {1} (\bibinfo {year} {2018})}\BibitemShut {NoStop}%
\bibitem [{\citenamefont {Kornev}\ \emph {et~al.}(2020)\citenamefont {Kornev}, \citenamefont {Kolotinskiy},\ and\ \citenamefont {Mukhanov}}]{kornevBiSQUIDDesignApplications2020}%
  \BibitemOpen
  \bibfield  {author} {\bibinfo {author} {\bibfnamefont {V.~K.}\ \bibnamefont {Kornev}}, \bibinfo {author} {\bibfnamefont {N.~V.}\ \bibnamefont {Kolotinskiy}},\ and\ \bibinfo {author} {\bibfnamefont {O.~A.}\ \bibnamefont {Mukhanov}},\ }\bibfield  {title} {{\selectlanguage {en}\bibinfo {title} {Bi-{SQUID}: design for applications}},\ }\href {https://doi.org/10.1088/1361-6668/aba541} {\bibfield  {journal} {\bibinfo  {journal} {Supercond. Sci. Technol.}\ }\textbf {\bibinfo {volume} {33}},\ \bibinfo {pages} {113001} (\bibinfo {year} {2020})}\BibitemShut {NoStop}%
\bibitem [{\citenamefont {Kornev}\ \emph {et~al.}(2011)\citenamefont {Kornev}, \citenamefont {Soloviev}, \citenamefont {Klenov}, \citenamefont {Sharafiev},\ and\ \citenamefont {Mukhanov}}]{5672560}%
  \BibitemOpen
  \bibfield  {author} {\bibinfo {author} {\bibfnamefont {V.~K.}\ \bibnamefont {Kornev}}, \bibinfo {author} {\bibfnamefont {I.~I.}\ \bibnamefont {Soloviev}}, \bibinfo {author} {\bibfnamefont {N.~V.}\ \bibnamefont {Klenov}}, \bibinfo {author} {\bibfnamefont {A.~V.}\ \bibnamefont {Sharafiev}},\ and\ \bibinfo {author} {\bibfnamefont {O.~A.}\ \bibnamefont {Mukhanov}},\ }\bibfield  {title} {\bibinfo {title} {Linear bi-squid arrays for electrically small antennas},\ }\href {https://doi.org/10.1109/TASC.2010.2091711} {\bibfield  {journal} {\bibinfo  {journal} {IEEE Trans. Appl. Supercond.}\ }\textbf {\bibinfo {volume} {21}},\ \bibinfo {pages} {713} (\bibinfo {year} {2011})}\BibitemShut {NoStop}%
\bibitem [{\citenamefont {De~Simoni}\ \emph {et~al.}(2022)\citenamefont {De~Simoni}, \citenamefont {Cassola}, \citenamefont {Ligato}, \citenamefont {Tettamanzi},\ and\ \citenamefont {Giazotto}}]{PhysRevApplied.18.014073}%
  \BibitemOpen
  \bibfield  {author} {\bibinfo {author} {\bibfnamefont {G.}~\bibnamefont {De~Simoni}}, \bibinfo {author} {\bibfnamefont {L.}~\bibnamefont {Cassola}}, \bibinfo {author} {\bibfnamefont {N.}~\bibnamefont {Ligato}}, \bibinfo {author} {\bibfnamefont {G.~C.}\ \bibnamefont {Tettamanzi}},\ and\ \bibinfo {author} {\bibfnamefont {F.}~\bibnamefont {Giazotto}},\ }\bibfield  {title} {\bibinfo {title} {Ultrahigh linearity of the magnetic-flux-to-voltage response of proximity-based mesoscopic bi-squids},\ }\href {https://doi.org/10.1103/PhysRevApplied.18.014073} {\bibfield  {journal} {\bibinfo  {journal} {Phys. Rev. Appl.}\ }\textbf {\bibinfo {volume} {18}},\ \bibinfo {pages} {014073} (\bibinfo {year} {2022})}\BibitemShut {NoStop}%
\bibitem [{\citenamefont {Usadel}(1970)}]{PhysRevLett.25.507}%
  \BibitemOpen
  \bibfield  {author} {\bibinfo {author} {\bibfnamefont {K.~D.}\ \bibnamefont {Usadel}},\ }\bibfield  {title} {\bibinfo {title} {Generalized diffusion equation for superconducting alloys},\ }\href {https://doi.org/10.1103/PhysRevLett.25.507} {\bibfield  {journal} {\bibinfo  {journal} {Phys. Rev. Lett.}\ }\textbf {\bibinfo {volume} {25}},\ \bibinfo {pages} {507} (\bibinfo {year} {1970})}\BibitemShut {NoStop}%
\bibitem [{\citenamefont {Likharev}(1979)}]{likharevSuperconductingWeakLinks1979}%
  \BibitemOpen
  \bibfield  {author} {\bibinfo {author} {\bibfnamefont {K.~K.}\ \bibnamefont {Likharev}},\ }\bibfield  {title} {{\selectlanguage {en}\bibinfo {title} {Superconducting weak links}},\ }\href {https://doi.org/10.1103/RevModPhys.51.101} {\bibfield  {journal} {\bibinfo  {journal} {Rev. Mod. Phys.}\ }\textbf {\bibinfo {volume} {51}},\ \bibinfo {pages} {101} (\bibinfo {year} {1979})},\ \bibinfo {note} {number: 1}\BibitemShut {NoStop}%
\bibitem [{\citenamefont {Golubov}\ \emph {et~al.}(2004)\citenamefont {Golubov}, \citenamefont {Kupriyanov},\ and\ \citenamefont {Il’ichev}}]{golubovCurrentphaseRelationJosephson2004}%
  \BibitemOpen
  \bibfield  {author} {\bibinfo {author} {\bibfnamefont {A.~A.}\ \bibnamefont {Golubov}}, \bibinfo {author} {\bibfnamefont {M.~Y.}\ \bibnamefont {Kupriyanov}},\ and\ \bibinfo {author} {\bibfnamefont {E.}~\bibnamefont {Il’ichev}},\ }\bibfield  {title} {{\selectlanguage {en}\bibinfo {title} {The current-phase relation in {Josephson} junctions}},\ }\href {https://doi.org/10.1103/RevModPhys.76.411} {\bibfield  {journal} {\bibinfo  {journal} {Rev. Mod. Phys.}\ }\textbf {\bibinfo {volume} {76}},\ \bibinfo {pages} {411} (\bibinfo {year} {2004})}\BibitemShut {NoStop}%
\bibitem [{\citenamefont {Dubos}\ \emph {et~al.}(2001)\citenamefont {Dubos}, \citenamefont {Courtois}, \citenamefont {Pannetier}, \citenamefont {Wilhelm}, \citenamefont {Zaikin},\ and\ \citenamefont {Sch\"on}}]{PhysRevB.63.064502}%
  \BibitemOpen
  \bibfield  {author} {\bibinfo {author} {\bibfnamefont {P.}~\bibnamefont {Dubos}}, \bibinfo {author} {\bibfnamefont {H.}~\bibnamefont {Courtois}}, \bibinfo {author} {\bibfnamefont {B.}~\bibnamefont {Pannetier}}, \bibinfo {author} {\bibfnamefont {F.~K.}\ \bibnamefont {Wilhelm}}, \bibinfo {author} {\bibfnamefont {A.~D.}\ \bibnamefont {Zaikin}},\ and\ \bibinfo {author} {\bibfnamefont {G.}~\bibnamefont {Sch\"on}},\ }\bibfield  {title} {\bibinfo {title} {Josephson critical current in a long mesoscopic s-n-s junction},\ }\href {https://doi.org/10.1103/PhysRevB.63.064502} {\bibfield  {journal} {\bibinfo  {journal} {Phys. Rev. B}\ }\textbf {\bibinfo {volume} {63}},\ \bibinfo {pages} {064502} (\bibinfo {year} {2001})}\BibitemShut {NoStop}%
\bibitem [{\citenamefont {Courtois}\ \emph {et~al.}(2008)\citenamefont {Courtois}, \citenamefont {Meschke}, \citenamefont {Peltonen},\ and\ \citenamefont {Pekola}}]{PhysRevLett.101.067002}%
  \BibitemOpen
  \bibfield  {author} {\bibinfo {author} {\bibfnamefont {H.}~\bibnamefont {Courtois}}, \bibinfo {author} {\bibfnamefont {M.}~\bibnamefont {Meschke}}, \bibinfo {author} {\bibfnamefont {J.~T.}\ \bibnamefont {Peltonen}},\ and\ \bibinfo {author} {\bibfnamefont {J.~P.}\ \bibnamefont {Pekola}},\ }\bibfield  {title} {\bibinfo {title} {Origin of hysteresis in a proximity josephson junction},\ }\href {https://doi.org/10.1103/PhysRevLett.101.067002} {\bibfield  {journal} {\bibinfo  {journal} {Phys. Rev. Lett.}\ }\textbf {\bibinfo {volume} {101}},\ \bibinfo {pages} {067002} (\bibinfo {year} {2008})}\BibitemShut {NoStop}%
\bibitem [{\citenamefont {Angers}\ \emph {et~al.}(2008)\citenamefont {Angers}, \citenamefont {Chiodi}, \citenamefont {Montambaux}, \citenamefont {Ferrier}, \citenamefont {Gu\'eron}, \citenamefont {Bouchiat},\ and\ \citenamefont {Cuevas}}]{PhysRevB.77.165408}%
  \BibitemOpen
  \bibfield  {author} {\bibinfo {author} {\bibfnamefont {L.}~\bibnamefont {Angers}}, \bibinfo {author} {\bibfnamefont {F.}~\bibnamefont {Chiodi}}, \bibinfo {author} {\bibfnamefont {G.}~\bibnamefont {Montambaux}}, \bibinfo {author} {\bibfnamefont {M.}~\bibnamefont {Ferrier}}, \bibinfo {author} {\bibfnamefont {S.}~\bibnamefont {Gu\'eron}}, \bibinfo {author} {\bibfnamefont {H.}~\bibnamefont {Bouchiat}},\ and\ \bibinfo {author} {\bibfnamefont {J.~C.}\ \bibnamefont {Cuevas}},\ }\bibfield  {title} {\bibinfo {title} {Proximity dc squids in the long-junction limit},\ }\href {https://doi.org/10.1103/PhysRevB.77.165408} {\bibfield  {journal} {\bibinfo  {journal} {Phys. Rev. B}\ }\textbf {\bibinfo {volume} {77}},\ \bibinfo {pages} {165408} (\bibinfo {year} {2008})}\BibitemShut {NoStop}%
\bibitem [{\citenamefont {Kong}\ \emph {et~al.}(2024)\citenamefont {Kong}, \citenamefont {Cruddas}, \citenamefont {Marenkovic}, \citenamefont {Tang}, \citenamefont {De~Simoni}, \citenamefont {Giazotto},\ and\ \citenamefont {Tettamanzi}}]{Kong_2024}%
  \BibitemOpen
  \bibfield  {author} {\bibinfo {author} {\bibfnamefont {T.~X.}\ \bibnamefont {Kong}}, \bibinfo {author} {\bibfnamefont {J.}~\bibnamefont {Cruddas}}, \bibinfo {author} {\bibfnamefont {J.}~\bibnamefont {Marenkovic}}, \bibinfo {author} {\bibfnamefont {W.}~\bibnamefont {Tang}}, \bibinfo {author} {\bibfnamefont {G.}~\bibnamefont {De~Simoni}}, \bibinfo {author} {\bibfnamefont {F.}~\bibnamefont {Giazotto}},\ and\ \bibinfo {author} {\bibfnamefont {G.~C.}\ \bibnamefont {Tettamanzi}},\ }\bibfield  {title} {\bibinfo {title} {Circuit-theoretic phenomenological model of an electrostatic gate-controlled bi-squid},\ }\href {https://doi.org/10.1088/1361-6668/ad813f} {\bibfield  {journal} {\bibinfo  {journal} {Superconductor Science and Technology}\ }\textbf {\bibinfo {volume} {37}},\ \bibinfo {pages} {115014} (\bibinfo {year} {2024})}\BibitemShut {NoStop}%
\bibitem [{\citenamefont {Ruf}\ \emph {et~al.}(2024)\citenamefont {Ruf}, \citenamefont {Puglia}, \citenamefont {Elalaily}, \citenamefont {De~Simoni}, \citenamefont {Joint}, \citenamefont {Berke}, \citenamefont {Koch}, \citenamefont {Iorio}, \citenamefont {Khorshidian}, \citenamefont {Makk} \emph {et~al.}}]{ruf2024gate}%
  \BibitemOpen
  \bibfield  {author} {\bibinfo {author} {\bibfnamefont {L.}~\bibnamefont {Ruf}}, \bibinfo {author} {\bibfnamefont {C.}~\bibnamefont {Puglia}}, \bibinfo {author} {\bibfnamefont {T.}~\bibnamefont {Elalaily}}, \bibinfo {author} {\bibfnamefont {G.}~\bibnamefont {De~Simoni}}, \bibinfo {author} {\bibfnamefont {F.}~\bibnamefont {Joint}}, \bibinfo {author} {\bibfnamefont {M.}~\bibnamefont {Berke}}, \bibinfo {author} {\bibfnamefont {J.}~\bibnamefont {Koch}}, \bibinfo {author} {\bibfnamefont {A.}~\bibnamefont {Iorio}}, \bibinfo {author} {\bibfnamefont {S.}~\bibnamefont {Khorshidian}}, \bibinfo {author} {\bibfnamefont {P.}~\bibnamefont {Makk}}, \emph {et~al.},\ }\bibfield  {title} {\bibinfo {title} {Gate control of superconducting current: Mechanisms, parameters, and technological potential},\ }\href@noop {} {\bibfield  {journal} {\bibinfo  {journal} {Applied Physics Reviews}\ }\textbf {\bibinfo {volume} {11}} (\bibinfo {year} {2024})}\BibitemShut {NoStop}%
\bibitem [{\citenamefont {De~Simoni}\ \emph {et~al.}(2019)\citenamefont {De~Simoni}, \citenamefont {Paolucci}, \citenamefont {Puglia},\ and\ \citenamefont {Giazotto}}]{desimoniJosephsonFieldEffectTransistors2019}%
  \BibitemOpen
  \bibfield  {author} {\bibinfo {author} {\bibfnamefont {G.}~\bibnamefont {De~Simoni}}, \bibinfo {author} {\bibfnamefont {F.}~\bibnamefont {Paolucci}}, \bibinfo {author} {\bibfnamefont {C.}~\bibnamefont {Puglia}},\ and\ \bibinfo {author} {\bibfnamefont {F.}~\bibnamefont {Giazotto}},\ }\bibfield  {title} {\bibinfo {title} {Josephson {Field}-{Effect} {Transistors} {Based} on {All}-{Metallic} {Al}/{Cu}/{Al} {Proximity} {Nanojunctions}},\ }\href {https://doi.org/10.1021/acsnano.9b02209} {\bibfield  {journal} {\bibinfo  {journal} {ACS Nano}\ }\textbf {\bibinfo {volume} {13}},\ \bibinfo {pages} {7871} (\bibinfo {year} {2019})},\ \bibinfo {note} {publisher: American Chemical Society}\BibitemShut {NoStop}%
\bibitem [{\citenamefont {De~Simoni}\ \emph {et~al.}(2021)\citenamefont {De~Simoni}, \citenamefont {Battisti}, \citenamefont {Ligato}, \citenamefont {Mercaldo}, \citenamefont {Cuoco},\ and\ \citenamefont {Giazotto}}]{desimoniGateControlCurrent2021a}%
  \BibitemOpen
  \bibfield  {author} {\bibinfo {author} {\bibfnamefont {G.}~\bibnamefont {De~Simoni}}, \bibinfo {author} {\bibfnamefont {S.}~\bibnamefont {Battisti}}, \bibinfo {author} {\bibfnamefont {N.}~\bibnamefont {Ligato}}, \bibinfo {author} {\bibfnamefont {M.~T.}\ \bibnamefont {Mercaldo}}, \bibinfo {author} {\bibfnamefont {M.}~\bibnamefont {Cuoco}},\ and\ \bibinfo {author} {\bibfnamefont {F.}~\bibnamefont {Giazotto}},\ }\bibfield  {title} {\bibinfo {title} {Gate {Control} of the {Current}–{Flux} {Relation} of a {Josephson} {Quantum} {Interferometer} {Based} on {Proximitized} {Metallic} {Nanojuntions}},\ }\href {https://doi.org/10.1021/acsaelm.1c00508} {\bibfield  {journal} {\bibinfo  {journal} {ACS Appl. Electron. Mater.}\ }\textbf {\bibinfo {volume} {3}},\ \bibinfo {pages} {3927} (\bibinfo {year} {2021})},\ \bibinfo {note} {publisher: American Chemical Society}\BibitemShut {NoStop}%
\bibitem [{\citenamefont {Bours}\ \emph {et~al.}(2020)\citenamefont {Bours}, \citenamefont {Mercaldo}, \citenamefont {Cuoco}, \citenamefont {Strambini},\ and\ \citenamefont {Giazotto}}]{bours2020unveiling}%
  \BibitemOpen
  \bibfield  {author} {\bibinfo {author} {\bibfnamefont {L.}~\bibnamefont {Bours}}, \bibinfo {author} {\bibfnamefont {M.~T.}\ \bibnamefont {Mercaldo}}, \bibinfo {author} {\bibfnamefont {M.}~\bibnamefont {Cuoco}}, \bibinfo {author} {\bibfnamefont {E.}~\bibnamefont {Strambini}},\ and\ \bibinfo {author} {\bibfnamefont {F.}~\bibnamefont {Giazotto}},\ }\bibfield  {title} {\bibinfo {title} {Unveiling mechanisms of electric field effects on superconductors by a magnetic field response},\ }\href@noop {} {\bibfield  {journal} {\bibinfo  {journal} {Physical Review Research}\ }\textbf {\bibinfo {volume} {2}},\ \bibinfo {pages} {033353} (\bibinfo {year} {2020})}\BibitemShut {NoStop}%
\bibitem [{\citenamefont {McCaughan}\ and\ \citenamefont {Berggren}(2014)}]{mccaughan_superconducting-nanowire_2014}%
  \BibitemOpen
  \bibfield  {author} {\bibinfo {author} {\bibfnamefont {A.~N.}\ \bibnamefont {McCaughan}}\ and\ \bibinfo {author} {\bibfnamefont {K.~K.}\ \bibnamefont {Berggren}},\ }\bibfield  {title} {{\selectlanguage {en}\bibinfo {title} {A {Superconducting}-{Nanowire} {Three}-{Terminal} {Electrothermal} {Device}}},\ }\href {https://doi.org/10.1021/nl502629x} {\bibfield  {journal} {\bibinfo  {journal} {Nano Lett.}\ }\textbf {\bibinfo {volume} {14}},\ \bibinfo {pages} {5748} (\bibinfo {year} {2014})}\BibitemShut {NoStop}%
\bibitem [{\citenamefont {Baghdadi}\ \emph {et~al.}(2020)\citenamefont {Baghdadi}, \citenamefont {Allmaras}, \citenamefont {Butters}, \citenamefont {Dane}, \citenamefont {Iqbal}, \citenamefont {McCaughan}, \citenamefont {Toomey}, \citenamefont {Zhao}, \citenamefont {Kozorezov},\ and\ \citenamefont {Berggren}}]{baghdadi_multilayered_2020}%
  \BibitemOpen
  \bibfield  {author} {\bibinfo {author} {\bibfnamefont {R.}~\bibnamefont {Baghdadi}}, \bibinfo {author} {\bibfnamefont {J.~P.}\ \bibnamefont {Allmaras}}, \bibinfo {author} {\bibfnamefont {B.~A.}\ \bibnamefont {Butters}}, \bibinfo {author} {\bibfnamefont {A.~E.}\ \bibnamefont {Dane}}, \bibinfo {author} {\bibfnamefont {S.}~\bibnamefont {Iqbal}}, \bibinfo {author} {\bibfnamefont {A.~N.}\ \bibnamefont {McCaughan}}, \bibinfo {author} {\bibfnamefont {E.~A.}\ \bibnamefont {Toomey}}, \bibinfo {author} {\bibfnamefont {Q.-Y.}\ \bibnamefont {Zhao}}, \bibinfo {author} {\bibfnamefont {A.~G.}\ \bibnamefont {Kozorezov}},\ and\ \bibinfo {author} {\bibfnamefont {K.~K.}\ \bibnamefont {Berggren}},\ }\bibfield  {title} {{\selectlanguage {en}\bibinfo {title} {Multilayered {Heater} {Nanocryotron}: {A} {Superconducting}-{Nanowire}-{Based} {Thermal} {Switch}}},\ }\href {https://doi.org/10.1103/PhysRevApplied.14.054011} {\bibfield  {journal} {\bibinfo  {journal} {Phys. Rev. Appl.}\ }\textbf {\bibinfo {volume} {14}},\ \bibinfo
  {pages} {054011} (\bibinfo {year} {2020})}\BibitemShut {NoStop}%
\bibitem [{\citenamefont {Buzzi}\ \emph {et~al.}(2023)\citenamefont {Buzzi}, \citenamefont {Castellani}, \citenamefont {Foster}, \citenamefont {Medeiros}, \citenamefont {Colangelo},\ and\ \citenamefont {Berggren}}]{buzzi_nanocryotron_2023}%
  \BibitemOpen
  \bibfield  {author} {\bibinfo {author} {\bibfnamefont {A.}~\bibnamefont {Buzzi}}, \bibinfo {author} {\bibfnamefont {M.}~\bibnamefont {Castellani}}, \bibinfo {author} {\bibfnamefont {R.~A.}\ \bibnamefont {Foster}}, \bibinfo {author} {\bibfnamefont {O.}~\bibnamefont {Medeiros}}, \bibinfo {author} {\bibfnamefont {M.}~\bibnamefont {Colangelo}},\ and\ \bibinfo {author} {\bibfnamefont {K.~K.}\ \bibnamefont {Berggren}},\ }\bibfield  {title} {{\selectlanguage {en}\bibinfo {title} {A nanocryotron memory and logic family}},\ }\href {https://doi.org/10.1063/5.0144686} {\bibfield  {journal} {\bibinfo  {journal} {Appl. Phys. Lett.}\ }\textbf {\bibinfo {volume} {122}},\ \bibinfo {pages} {142601} (\bibinfo {year} {2023})}\BibitemShut {NoStop}%
\bibitem [{\citenamefont {Savin}\ \emph {et~al.}(2004)\citenamefont {Savin}, \citenamefont {Pekola}, \citenamefont {Flyktman}, \citenamefont {Anthore},\ and\ \citenamefont {Giazotto}}]{savin2004cold}%
  \BibitemOpen
  \bibfield  {author} {\bibinfo {author} {\bibfnamefont {A.}~\bibnamefont {Savin}}, \bibinfo {author} {\bibfnamefont {J.~P.}\ \bibnamefont {Pekola}}, \bibinfo {author} {\bibfnamefont {J.}~\bibnamefont {Flyktman}}, \bibinfo {author} {\bibfnamefont {A.}~\bibnamefont {Anthore}},\ and\ \bibinfo {author} {\bibfnamefont {F.}~\bibnamefont {Giazotto}},\ }\bibfield  {title} {\bibinfo {title} {Cold electron josephson transistor},\ }\href@noop {} {\bibfield  {journal} {\bibinfo  {journal} {Applied physics letters}\ }\textbf {\bibinfo {volume} {84}},\ \bibinfo {pages} {4179} (\bibinfo {year} {2004})}\BibitemShut {NoStop}%
\bibitem [{\citenamefont {Trupiano}\ \emph {et~al.}(2025{\natexlab{a}})\citenamefont {Trupiano}, \citenamefont {De~Simoni},\ and\ \citenamefont {Giazotto}}]{PhysRevApplied.23.014046}%
  \BibitemOpen
  \bibfield  {author} {\bibinfo {author} {\bibfnamefont {G.}~\bibnamefont {Trupiano}}, \bibinfo {author} {\bibfnamefont {G.}~\bibnamefont {De~Simoni}},\ and\ \bibinfo {author} {\bibfnamefont {F.}~\bibnamefont {Giazotto}},\ }\bibfield  {title} {\bibinfo {title} {Quasiparticle-injection superconducting microwave relaxation oscillator},\ }\href {https://doi.org/10.1103/PhysRevApplied.23.014046} {\bibfield  {journal} {\bibinfo  {journal} {Phys. Rev. Appl.}\ }\textbf {\bibinfo {volume} {23}},\ \bibinfo {pages} {014046} (\bibinfo {year} {2025}{\natexlab{a}})}\BibitemShut {NoStop}%
\bibitem [{\citenamefont {Tirelli}\ \emph {et~al.}(2008)\citenamefont {Tirelli}, \citenamefont {Savin}, \citenamefont {Garcia}, \citenamefont {Pekola}, \citenamefont {Beltram},\ and\ \citenamefont {Giazotto}}]{tirelli2008manipulation}%
  \BibitemOpen
  \bibfield  {author} {\bibinfo {author} {\bibfnamefont {S.}~\bibnamefont {Tirelli}}, \bibinfo {author} {\bibfnamefont {A.}~\bibnamefont {Savin}}, \bibinfo {author} {\bibfnamefont {C.~P.}\ \bibnamefont {Garcia}}, \bibinfo {author} {\bibfnamefont {J.~P.}\ \bibnamefont {Pekola}}, \bibinfo {author} {\bibfnamefont {F.}~\bibnamefont {Beltram}},\ and\ \bibinfo {author} {\bibfnamefont {F.}~\bibnamefont {Giazotto}},\ }\bibfield  {title} {\bibinfo {title} {Manipulation and generation of supercurrent in out-of-equilibrium<? format?> josephson tunnel nanojunctions},\ }\href@noop {} {\bibfield  {journal} {\bibinfo  {journal} {Physical review letters}\ }\textbf {\bibinfo {volume} {101}},\ \bibinfo {pages} {077004} (\bibinfo {year} {2008})}\BibitemShut {NoStop}%
\bibitem [{\citenamefont {Trupiano}\ \emph {et~al.}(2025{\natexlab{b}})\citenamefont {Trupiano}, \citenamefont {Simoni},\ and\ \citenamefont {Giazotto}}]{trupiano2025thermallymodulatedsinistrasconductance}%
  \BibitemOpen
  \bibfield  {author} {\bibinfo {author} {\bibfnamefont {G.}~\bibnamefont {Trupiano}}, \bibinfo {author} {\bibfnamefont {G.~D.}\ \bibnamefont {Simoni}},\ and\ \bibinfo {author} {\bibfnamefont {F.}~\bibnamefont {Giazotto}},\ }\href {https://arxiv.org/abs/2505.21341} {\bibinfo {title} {A thermally modulated sinis trasconductance amplifier}} (\bibinfo {year} {2025}{\natexlab{b}}),\ \Eprint {https://arxiv.org/abs/2505.21341} {arXiv:2505.21341 [cond-mat.supr-con]} \BibitemShut {NoStop}%
\bibitem [{\citenamefont {Ronzani}\ \emph {et~al.}(2013)\citenamefont {Ronzani}, \citenamefont {Baillergeau}, \citenamefont {Altimiras},\ and\ \citenamefont {Giazotto}}]{10.1063/1.4817013}%
  \BibitemOpen
  \bibfield  {author} {\bibinfo {author} {\bibfnamefont {A.}~\bibnamefont {Ronzani}}, \bibinfo {author} {\bibfnamefont {M.}~\bibnamefont {Baillergeau}}, \bibinfo {author} {\bibfnamefont {C.}~\bibnamefont {Altimiras}},\ and\ \bibinfo {author} {\bibfnamefont {F.}~\bibnamefont {Giazotto}},\ }\bibfield  {title} {\bibinfo {title} {Micro-superconducting quantum interference devices based on {V}/{Cu}/{V} {Josephson} nanojunctions},\ }\href {https://doi.org/10.1063/1.4817013} {\bibfield  {journal} {\bibinfo  {journal} {Appl. Phys. Lett.}\ }\textbf {\bibinfo {volume} {103}},\ \bibinfo {pages} {052603} (\bibinfo {year} {2013})}\BibitemShut {NoStop}%
\end{thebibliography}
\end{document}